\documentclass[%
superscriptaddress,
preprint,
 amsmath,amssymb,
 aps,
]{revtex4-2}

\usepackage{graphicx}
\graphicspath{{Figures/}}
\usepackage{dcolumn}
\usepackage{bm}

\usepackage{color}
\usepackage[dvipsnames]{xcolor}

\begin{document}

\setlength{\abovedisplayskip}{3pt}
\setlength{\belowdisplayskip}{3pt}

\title{Finite amplitude wave propagation through bubbly fluids}

\author{Siew-Wan Ohl}
\affiliation{Department of Soft Matter, Institute of Physics \& Faculty of Natural Sciences, University of Magdeburg, Universitätplatz 2, 39108 Magdeburg, Germany.}
\affiliation{Research Campus STIMULATE, University of Magdeburg, Otto-Hahn-Straße 2, 39106 Magdeburg, Germany.}
\author{Juan Manuel Rosselló}
\affiliation{Department of Soft Matter, Institute of Physics \& Faculty of Natural Sciences, University of Magdeburg, Universitätplatz 2, 39108 Magdeburg, Germany.}
\author{Daniel Fuster}
\affiliation{Institut Jean Le Rond D'Alembert UPMC, Tour 55-65, 75005 Paris, France.}
\author{Claus-Dieter Ohl}
\affiliation{Department of Soft Matter, Institute of Physics \& Faculty of Natural Sciences, University of Magdeburg, Universitätplatz 2, 39108 Magdeburg, Germany.}

\date{\today}

\begin{abstract}
The existence of only a few bubbles could drastically reduce the acoustic wave speed in a liquid. Wood’s equation models the linear sound speed, while the speed of an ideal shock waves is derived as a function of the pressure ratio across the shock. The common finite amplitude waves lie, however, in between these limits. We show that in a bubbly medium, the high frequency components of finite amplitude waves are attenuated and dissipate quickly, but a low frequency part remains. This wave is then transmitted by the collapse of the bubbles and its speed decreases with increasing void fraction. We demonstrate that the linear and the shock wave regimes can be smoothly connected through a Mach number based on the collapse velocity of the bubbles.
\end{abstract}

\maketitle

\section{Introduction}
Since the observation of damping of sound in frothy liquid in 1910~\citep{mallock1910damping}, the physics of sound propagation in bubbly liquid has attracted many research works. The simplest models make the assumptions that the bubbly liquid is a continuum, the sound wave has a small amplitude, and the bubbles are uniformly distributed~\citep{minnaert1933xvi,carstensen1947propagation,meyer1953akustischen,Wijngaarden1968}.~\citet{Wijngaarden1968}~showed that a void fraction of 0.3\% reduces the speed of sound to about 200 m/s, from 1500 m/s in the pure liquid. 

Wave velocity and attenuation depend on frequency in a complex way. Effective medium approaches estimate that the dispersion relation of pure gas bubbles~\citep{carstensen1947propagation,wijngaarden1972one,commander1989linear} and gas/vapor bubbles~\citep{fuster2015mass}~to be in the linear regime. At low frequencies, dispersion effects are negligible, and the effective speed approaches an asymptotic value known as Wood's limit~\citep{wood1941textbook}: 

\begin{equation}
\frac{1}{c^2}=\frac{(1-\alpha)^2}{c_l^2}+\frac{\rho_l \alpha(1-\alpha)}{\kappa\,p}\quad ,
\label{Eq:Wood}
\end{equation}
where $c_l$ and $c_g$ are the speed of sound in the liquid and gas respectively, $\alpha$ is the void fraction, and $\kappa$ the ratio of specific heats~\citep{Wijngaarden2007}.~\citet{silberman1957sound}, among others, have measured the sound velocity in a standing wave tube filled with a mixture of bubbles and water, and verified Wood's derivation.

For the strongly nonlinear regime where the small amplitude assumption is not valid,~\citet{campbell1958shock} connected Eqn.~(\ref{Eq:Wood}) with the propagation speed of a shock wave, $U$. They used the shock relations for continuity, momentum, and energy in a bubbly liquid and obtained
\begin{equation}
 U^2=c^2\left(\frac{p_1}{p_0}-1\right)\left(1-\left(\frac{p_0}{p_1}\right)^{1/\kappa}\right)^{-1}
\quad ,
\label{Eq:C&P}
\end{equation}
with $p_0$ and $p_1$ being the pressures in the bubbly liquid ahead and behind the shock wave, respectively~\citep{Wijngaarden2007}. This expression for $\kappa=1$, i.e.~$U^2 = (p_1/p_0) c^2$, was confirmed by \citet{campbell1958shock} with a vertical shock tube experiment where small gas bubbles and void fractions between~$\alpha=5\%$~and~$\alpha=30\%$~were used. It is noted that Eqn. (\ref{Eq:C&P}) is only valid for sufficiently high void fractions and thus $c \ll c_l$, for which
the effective sound speed of Eqn.~\ref{Eq:Wood}~no longer depends on $c_l$
$$c \approx \sqrt{\frac{\kappa p_0}{\rho_l \, \alpha (1 - \alpha).}}$$ 
Interestingly, this theory predicts that the effective speed of sound of the medium is proportional to $\sqrt{p_1/p_0}$.

Between the linear and the strongly nonlinear regimes, models for dispersion and relaxation of weak and moderate-strength shock waves have been established and tested experimentally. Kameda et al.~\citep{kameda1996shock,kameda1998shock}~confirmed Wijngaarden's~\citep{Wijngaarden1968}~bubble-liquid mixture model with~$\Delta p = p_1-p_0$~up to $1\,$bar. \citet{ando2011numerical}~modeled bubble poly-dispersity's effect on shock front shape, and found that the broad bubble size distributions smooth out the profile. Therefore, the linear theory is found not to be suitable for predicting strong shock propagation speed.

In this paper, we study how finite amplitude waves propagate through bubbles in gelatin. Our research is inspired by Dear and Field's classic work on shock waves interaction with gas bubbles in hydrogels~\citep{dear1988gas,dear1988study}. Firstly, we use Direct Numerical Simulation to understand wave transmission across all acoustic regimes. After that, the experiments on compressive wave propagation through bubbles in gelatin gels are presented. We found that the experimental results are in agreement with DNS simulations on high-pressure finite amplitude wave regimes.

\section{Direct Numerical Simulation} 
The simulations are conducted with a multiphase compressible solver presented in~\citet{fuster2018all}. 
This solver for the Navier--Stokes equations in a Newtonian fluid has previously been applied for sub-harmonic emissions of bubbles oscillating in a tube \citep{fan2020optimal}.\\

\begin{figure}
\includegraphics[width=0.8\linewidth]{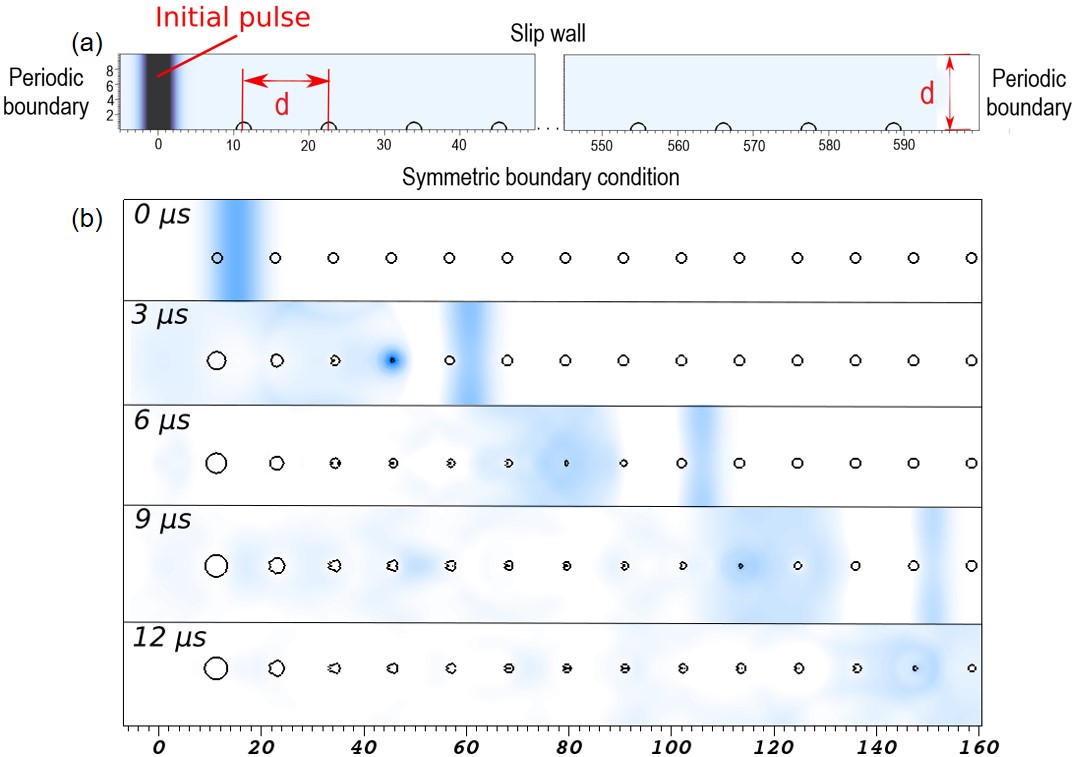} \setlength{\belowcaptionskip}{-5pt} 
\caption{\label{fig:DNS}(a) Domain and boundary conditions used for DNS simulations. (b) A series of snapshots of the pressure field and interface contour at an initial void fraction,~$\alpha = 9.2 \times 10^{-4}$ and $\Delta p/p_0=100$ with indicated timing.}
\end{figure}

We model an idealized system where a pressure pulse propagates through a mono-dispersed line of bubbles with size~$R_0=100\,\mu$m~in a viscous fluid (Figure~\ref{fig:DNS}(a)). We consider a periodic axisymmetric configuration where the bubbles are uniformly distributed along the axis of symmetry at a constant inter-bubble distance $d$. On the top domain boundary, slip boundary conditions are applied. The simulation domain covers the range $-d/2 \le z \le 52.5d$ and 0 $\le r \le d$ in the radial direction. A uniform grid of
size ${\Delta x}/{d}={1}/{128}$ is used to carry out the simulations, which makes in total 868,352 grid points. With this configuration, the effective void fraction is $\alpha=4/3(R_0/d)^3$.

A Gaussian pulse, denoted by equations
\begin{eqnarray}
&&p(z,t=0)  =  p_0 + \Delta p e^{-z^2/\lambda^2} \\
&& u_z(z,t=0)  =  \frac{p(z,0)-p_0}{\rho_l c_l}
\end{eqnarray}~is propagating in the simulation domain from left to right. The pulse shape is designed to mimic the finite amplitude wave profile in the experiment. The initial amplitude of the pressure pulse is left as a variable parameter, and the wavelength~$\lambda$~is set to~$\lambda=500~\mu$m.
The spectral content gaussian pulse considered
is also a gaussian function where the range of forcing frequencies excited is  $\frac{\omega}{\omega_0} \le 100$
with $\omega_0$ being the bubble natural resonant frequency. The gas inside the bubble is modeled as an ideal gas. The viscous fluid is assumed to be water-like but with a higher viscosity ($\nu_l=2.3\times10^{-4} \mathrm{m^2/s}$)(see supplementary file "Viscosity derivation" for value derivation), although this parameter 
has revealed not to be crucial in describing the physics in what follows.

Figure~\ref{fig:DNS}(b) depicts various snapshots of the pressure field and the bubble interface contour during the propagation of a wave with amplitude~$\Delta p/p_0=100$. The precursor wave, which is the high frequency component of the finite amplitude wave, propagates across the bubbly medium at approximately the speed of sound in the pure liquid, inducing an initial compression of the bubbles. The term ``high frequency'' here is referring to the signal frequencies which are higher than the bubble resonance frequency,~$\omega \gg \omega_0$. There is an increasing delay between the instant at which the precursor wave reaches a given bubble location and the instant at which the bubble volume becomes minimum.

\begin{figure}
\includegraphics[width=0.6\linewidth]{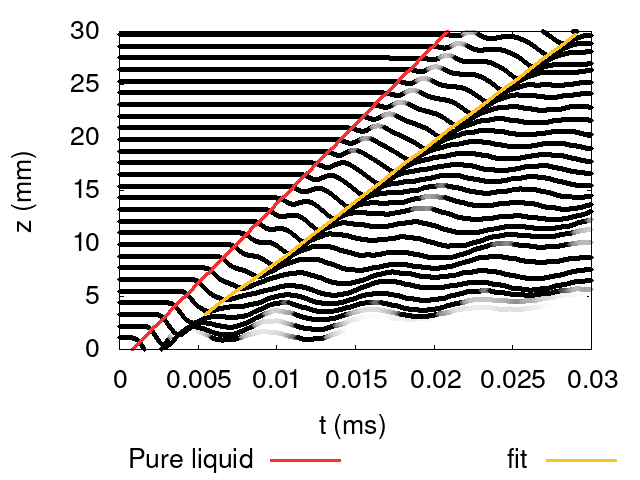} \setlength{\belowcaptionskip}{-5pt} 
\caption{\label{fig:DNSxtdiagram} Spatio-temporal representation of the evolution of the bubble volume
of the first 30 bubbles contained in the domain.
The red line represents the theoretical propagation of a
wave propagating at the pure liquid speed of sound.
The yellow line represents a line fitting the instant
of minimum radius of the various bubbles.}
\end{figure}
Figure \ref{fig:DNSxtdiagram} depicts the 
temporal evolution of the bubble volume of the first 30
bubbles contained in the domain in a spatio-temporal diagram. The figure reveals that 
we can fit a straight line going through
the points defined by the instant of minimum radius
as a function of the bubble position. The slope of this line defines a clear and unique effective propagation velocity at which the front associated to the collapse of bubbles propagates. This velocity cannot be directly associated with the velocity of the pressure pulse, where existing theories predict the
dispersion relation where the effective phase velocity depends on the frequency. More description and discussion of the wave propagation is given in the supplementary file "Additional Explanation and Simulation".

\begin{figure}
\includegraphics[width=0.45\linewidth]{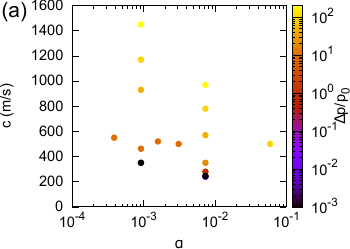} \\
\includegraphics[width=0.45\linewidth]{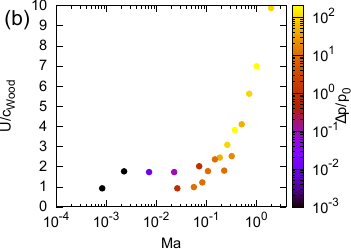} 
\includegraphics[width=0.45\linewidth]{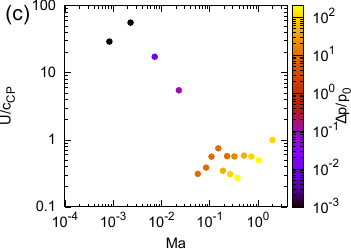} 
\caption{\label{fig:DNSplots}(a) Effective wave velocity for all values tested of~$\Delta p/p_0$ and void fraction. 
(b) Velocities rescaled with the Wood velocity as a function of the Mach number. (c) Velocities rescaled with the Campbell \& Pitcher velocity as a function of the Mach number. }
\end{figure}

To further examine the behavior of this effective wave, Figure~\ref{fig:DNSplots} shows the effect of the strength of the propagating compression wave, $\Delta p/p_0$, and the void fraction on the value of the effective velocity obtained. Defining the Mach number as the ratio between  the characteristic speed of the bubble collapse and the linear speed of sound in the bubbly liquid, $Ma=\frac{1}{c_{wood}} \sqrt{\Delta p/\rho_l}$,  it is possible to collapse well the dimensional values obtained for 
all the different conditions tested. The derivation of the Ma number is presented in detail in the Section~\ref{sec:discussion}. The velocities obtained for low amplitudes of the forcing pulse (low Ma) are independent of the Mach number (e.g. the forcing amplitude) and are predicted by the Wood theory. 
This points to the fact that the effective wave speed is controlled by the low frequency content of the signal, which have small effective attenuation in comparison with
the high frequency components.
Figure~\ref{fig:DNSplots} reveals that as
we approach values of the Mach number close to one, the Campbell \& Pitcher velocity serves to collapse the results for the various bubble concentration and pressure amplitudes tested.
The limitations of numerical simulations on the maximum values of the Mach number that can be reached or the differences of the initial pulse profile used in this work with respect to the shock
propagation problem considered in the theory may be behind 
the differences in the prefactor, but the results seem to point out
that, at least at first order, the nondimensional velocity
$U/U_{CP}$ tends to a value close to one
for $Ma > 0.2$.

\section{Experimental setup and observations} \label{sec:exp}
\begin{figure}
\includegraphics[width=0.5\linewidth]{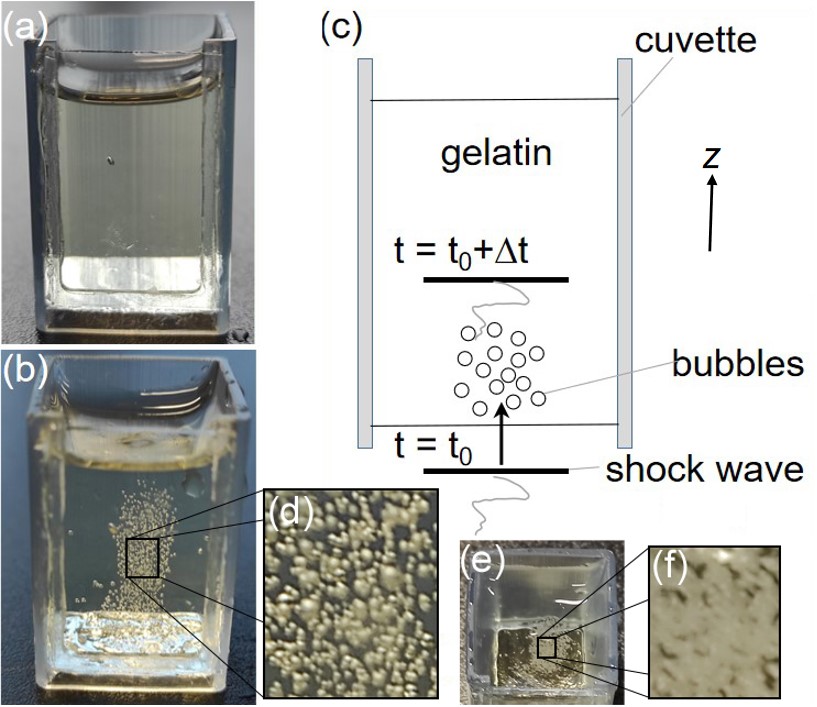}
\caption{\label{fig:setup}(a) Gelatin in a customised cuvette prior to the experiment. One face of the cuvette is replaced with glass and the bottom of the cuvette is removed. The dimension is 20.3$\times$20.3$\times$30.0~$\mathrm{mm^3}$. The center of the cuvette is placed at the focal volume, approximately 135 mm from the shock wave generator surface (where~$z$~= 0). For pressure measurement, a hydrophone is placed about 10 mm from the bottom of the cuvette and 2 mm left from the center. (b) The cuvette after 10 wave applications resulting in a central cloud of bubbles trapped in the gelatin. (c) The zoomed areas show the limits of the observation window of the high-speed videos (region of interest) viewed from the side. (e) The cuvette as viewed from the top and (f) the corresponding region of interest as observed from the top. It is noted that the camera is viewing from the side only.}
\end{figure}

The experiments are carried out using gelatin gel which is placed within a custom-made cuvette. This cuvette is cut from a square cross section extruded plastic tube (K.+C. Weiss GmbH) with dimensions 20.3$\times$20.3$\times$30.0~$\mathrm{mm^3}$. One of the plastic side wall of the cuvette is replaced with glass to facilitate good visualization of the bubbles (Figure~\ref{fig:setup}(a)). During the experiment the bottom and top of the cuvette is open, and the whole cuvette is submerged in the water bath, allowing both the top and bottom gelatin gel surfaces to be in contact with water. This arrangement allows for good acoustic coupling as the compressive wave propagates through the gel. 

The gelatin is prepared from industrial gelatin granules made from pork or beef skin (Gelatin 250 bloom, Yasin Gelatin, China). The gelatin granules are weighted and mixed with deionised water. A 4\% gelatin contains 4 grams of gelatin granules per 100 grams of the gelatin-water mixture. Gelatin gel of various concentrations are prepared (4\%, 6\%, and 8\%). The mixture is then heated under stirring on a hot plate to around $80^\circ\,$C. Once homogeneity is reached, the hot mixture is removed from the hot plate, and poured into a customised plastic cuvette with the bottom lid on. The cuvette with gelatin is placed in a refrigerator at 5~$^{\circ}\mathrm{C}$~for 72 hours. The long refrigeration time is needed because it is observed that if the gelatin gel has been stored for less than 48 hours, the bubbles created by the waves in the gelatin gel will dissolve over a time scale of a few seconds. However, if the gelatin sample has been kept for more than 72 hours, the gas bubbles grow slowly with time once nucleated. We speculate that the long storage time is needed to saturate the gelatin gel with air. The heating during preparation depleted the sample from its initial saturation concentration at room temperature. Once the sample is brought back to room temperature, the aged gelatin is supersaturated with air and can't equilibrate in the duration of the experiment. Thus a sample that has achieved room temperature is still supersaturated, causing the growth of gas bubbles against the elastic forces.

At least two hours before the experiment, the cuvette is taken out of the fridge and is allowed to return to room temperature. It is then placed in the focal volume of the shock wave generator in the water tank during the experiment. The shock wave generator is a modified medical lithotripter equipped with two layers of piezoelectric transducers. In this study, only the top piezo layer is used, and it is operated at $7\,$kV. The repeatability of the pressure signatures in water is high, for details see~\citet{Arora2005}. 

Initially the gelatin gel is clear and is absent of gas bubbles. Only after the first passage of the finite amplitude wave, gas bubbles are formed along the acoustic path in the center of the cuvette due to the tensile part of the wave. These bubbles do not dissolve and they persist in the gelatin. A waiting interval of 10 minutes is set between successive wave admission. During this time, the bubbles grow in size via diffusion of gas from the gel into the bubble, and thus the void fraction is increased. A photograph of a typical sample taken after 100 minutes (i.e. after 10 applied finite amplitude waves) is shown in Figure~\ref{fig:setup}(b). The experiment is then concluded for this sample. Although the bubbles are nucleated only in the focal region of the shock waves generator, within the observation window of the high speed camera (Figure~\ref{fig:setup}(d)), the bubbles are distributed homogeneously throughout the frame. Similarly from the top of the cuvette (Figure~\ref{fig:setup}(e)), the region of interest (Figure~\ref{fig:setup}(f)) can be seen to be filled with bubbles. The boundaries of the bubble cloud are not studied in this experiment.

\begin{figure}
\includegraphics[width=0.8\linewidth]{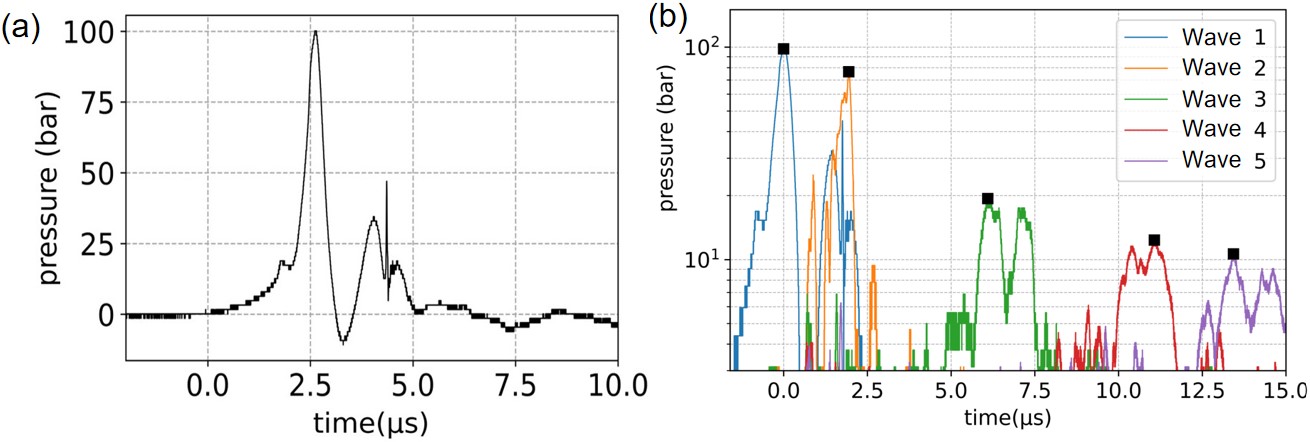}
\caption{\label{fig:pressuremeasured} (a) A typical measurement of the finite amplitude wave generated with the lithotripter. Maximum pressure of around 100 bar is measured in gelatin without bubbles. (b) The pressure peaks (black squares) as measured by the needle hydrophone placed in 4\% gelatin approximately 2 mm to the left of the focal volume, in a separate experiment where all other parameters remain the same. With successive waves, the wave arrival is delayed and its amplitude is reduced.}
\end{figure}

The pressure is measured with a needle hydrophone (diameter 1.1 mm, model Müller-Platte Nadelsonde) inserted into the gelatin cuvette about $2\,$mm sideways from the focus. The hydrophone is connected to an oscilloscope (DS1054Z, Rigol Technology Inc). The first bubbles are nucleated in the tensile region with approximately $-10\,$bar amplitude when the finite amplitude wave as shown in~\ref{fig:pressuremeasured}(a) passes through the gelatin gel. The bubble dynamics is captured with a high speed camera (HPV2, Shimadzu Inc.) together with a generic camera flash light at a framing rate of 1 million frames per second and a pixel resolution of 312 x 260 pixels. The timing of the devices is controlled with a digital delay generator (BNC 525, Berkeley Nucleonics). 

The slow down of the compression wave is observed in the hydrophone measurements in Figure~\ref{fig:pressuremeasured}(b). During the measurements, the position of the hydrophone in a $4\,$\% gelatin sample was fixed while consecutive finite amplitude waves at an interval of $10\,$minutes are applied. Wave 1 in Figure~\ref{fig:pressuremeasured}(b)~relates to the virgin sample, and the later waves are interacting with the bubbles. The time,~$t = 0$, corresponds to the moment when wave 1 is detected. The pressure peaks of subsequent waves are shifted to later times and are considerably reduced in amplitude from the initial $100\,$bar to $10\,$bar after only 5 finite amplitude waves. This observation suggests that the wave speed is a function of the void fraction (which increased over time as the bubble grows in size over the multiple application of the finite amplitude waves), as well as the wave amplitude since the wave is dissipated by the bubble oscillations.

\begin{figure}
\includegraphics[width=1.0\linewidth]{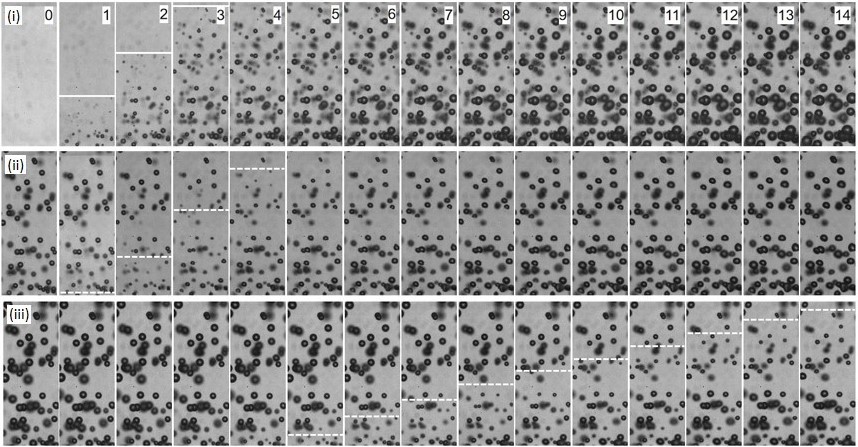} \setlength{\belowcaptionskip}{-5pt} 
\caption{\label{fig:ExpRes} Bubble nucleation and compression by a traveling finite amplitude wave in 4\% gelatin. The frame height is $4.974\,$mm and the time is indicated in microseconds. (i)$1^\mathrm{st}$ finite amplitude wave, (ii) $2^\mathrm{nd}$ wave, and (iii) $8^\mathrm{th}$ wave.}
\end{figure}

Figure~\ref{fig:ExpRes}(i)~depicts a high-speed recording of the first pressure pulse propagating in a virgin gelatin block (4\% by weight) from the bottom to the top in positive $z$-direction. The time between frames is $1\,\mu$s. Within the first 4 frames, the bubbles are nucleated over a distance of about $5\,$mm. The approximate location where the rarefaction wave starts is indicated with a solid white line. In the later frames, the bubbles grow to a larger size and collapse approximately $150\,\mu$s later (not shown here). In contrast to the bubbles observed in a liquid, these gas bubbles in gelatin do not quickly dissolve or rise out of the field of view. Once formed, they remain at their location. This is utilized by allowing the bubbles to grow in between wave applications for about 10 minutes through the diffusion of gasses from the gelatin~\cite{ando2019}, thus to increase the void fraction. This void fraction growth is repeated 10 times. After about 100 minutes, the experiment for one gelatin sample is concluded.

The frames obtained from the high-speed recording are analyzed with home-built algorithms~\citep{rossello2023}~to detect the bubbles, register their locations and shapes, and link them across frames to follow their dynamics. The wave speed was then obtained by performing a numerical fit on the position where the bubbles reach a minimum size over time using a linear model. Initially, there is no bubble in the gelatin gel (Figure~\ref{fig:ExpRes}(i)), and thus to get the speed for the first finite amplitude wave, we track the bubble nucleation as the tensile phase of the wave passes through. Details about the gelatin gel elasticity measurement and image processing are provided in the appendixes.

Figure~\ref{fig:ExpRes}(ii) shows the compressive wave that passes through the gel that contains gas bubbles from the first wave administration. Over time, we see an upwards propagating region of compressed bubbles, i.e. over $3\,\mu$s in the frames from $1\,\mu$s $\le t \le 4\,\mu$s. The approximate location of the minimum bubble volume is indicated with the dash horizontal line. Figure~\ref{fig:ExpRes}(iii)~shows the bubble dynamics induced by the 8$^\mathrm{th}$ finite amplitude wave. We find a considerable delay in the arrival of the compressive wave, which enters the field of view from below only at $t=5\,\mu$s. This later arrival time is confirmed by the considerably slower propagation of the wave front. Now the compressed bubbles are seen at the top of the frame in the right-most frame, i.e. $t=14\,\mu$s. The average speed by which the compression region propagates upwards can be very roughly estimated from the distance of $4.5\,$mm over $9\,\,\mu$s. This wave speed is only $\approx 500\,$m/s! If we compare this value with the speed of the tensile wave propagating through the virgin gelatin in Figure~\ref{fig:ExpRes}(i), the wave travels within $3\,\mu$s from the bottom to the top of the frame, a distance of $5\,$mm, i.e. at a speed of about $1600\,$m/s. The high-speed recordings are provided as supplementary materials.

\section{Discussion} \label{sec:discussion}
\begin{figure}
\includegraphics[width=0.8\linewidth]{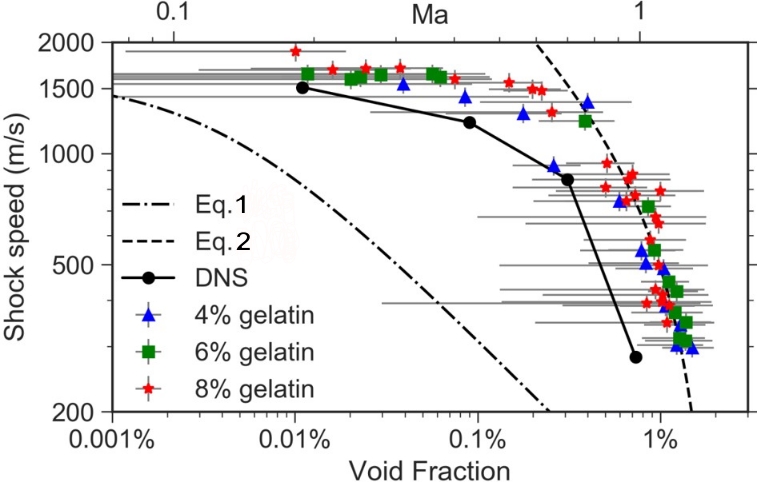} \setlength{\belowcaptionskip}{-5pt}
\caption{\label{fig:MainResult} Collected finite amplitude wave velocities for the 4\%, 6\%, and 8\% gelatin as a function of the void fraction. The graph is compared to the acoustic speed of sound, $c$, from Eqn.~\ref{Eq:Wood}, the shock velocity, $U$, from Eqn.~\ref{Eq:C&P}, and the results from DNS simulations. The upper horizontal scale is the Mach number, $Ma=c^{-1}(\Delta p/\rho)^{1/2}$.}
\end{figure}

Experimental results from 61 individual measurements of the compression wave speed in 6 gelatin samples with concentrations of $4\,$\%, $6\,$\%, and $8\,$\% by weight are shown in Figure~\ref{fig:MainResult}. The lowest velocity measured is $298\,$m/s at a void fraction of $1.5\,$\%. The wave speed is a function of the void fraction and the pressure amplitude. Thus, the measured pressure is fitted to a polynomial function of order 2 to obtain an approximated function of the pressure amplitude as a function of the void fraction. The dashed line in Figure~\ref{fig:MainResult} is obtained from Eqn.~\ref{Eq:C&P} using this fitted function for the pressure. Additionally, we plot in Figure~\ref{fig:MainResult} Wood's speed of sound Eqn.~\ref{Eq:Wood} as a dashed-dotted line. The latter predicts a much faster reduction of the propagation speed than measured. Clearly, important characteristics of the finite amplitude waves need to be accounted for. The expression of Campbell \& Pitcher (Eqn.~\ref{Eq:C&P}), however, is in reasonable agreement with the experimental results and is within measurement errors for sufficiently high void fractions starting from about $0.3\,$\%. For smaller void fractions, the measured pressure amplitude is above $70\,$bar. The combination of small void fraction and high-pressure ratio $p_1/p_0$ limits the applicability of Eqn.~\ref{Eq:C&P}. Figure~\ref{fig:MainResult} also compares the measured wave speed with the speed of bubble compression from the numerical simulations (DNS) (solid line). 

We use a Mach number to characterize the different wave regimes. By considering the importance of nonlinear wave propagation phenomena, the Mach number is quantified by the ratio between the characteristic flow velocity $U_c$ and the wave propagation velocity in the linear regime of the flow $c$, that is, the Mach number $Ma=U_c/c$. The characteristic flow velocity $U_c$ is derived from the average velocity for a bubble to collapse, which is $U_c=R_{max}/T_C$ where $R_{max}$ is the maximum bubble size and $T_C$ the collapse time. Using the approximate relationship of the Rayleigh (inertial) collapse time~$T_C\approx R_{max}\sqrt{\rho/\Delta p}$,~we obtain for the average speed of collapse $U_c=\sqrt{\Delta p/\rho}$. 

It is noted that the response of our system transits between the two limiting regimes, the linear regime~$Ma_{LF} \ll 1$, and the fully supersonic regime $Ma_{HF} \ge 1$, where Campbell and Pitcher's theory certainly applies.
This theory is indeed shown to provide a good first-order approximation to the influence of the bubble concentration and pressure amplitude on the evolution of the finite amplitude waves in the gelatin gel. This can be justified if we assume that the energy exchanges between the long wavelengths and the short wavelengths (or low and high frequencies) can be neglected at the leading order. These effects only appear as a second-order correction when $Ma_{LF} \, \ge \, 1$. Overall, the DNS provides a better agreement with the experimental data considering the simplifications involved. This agreement supports the explanation that with increasing void fraction, a low frequency wave with an approximately constant speed compresses the bubbles.

\section{Conclusion} 
We presented a Direct Numerical Simulation of a pressure pulse through a line of mono-dispersed bubbles. The bubbles cause the dispersion of the initial pulse into a high frequency component (the precursor wave), and a low frequency component which is propagated by the collapsing bubbles. The precursor wave, which travels at sound speed in pure liquid, vanishes quickly. Essentially, the collapsing bubbles act as a low-band filter that mainly propagates the low frequency component of the initial pulse. The increase in bubble size, and thus the void fraction, reduce both the amplitude and speed of the propagated wave.

Experimentally, we used high-speed photography to observe the slow-down of the compressive wave by following the bubble collapses. The gas bubbles are generated in the rarefaction part of the first lithotripter wave administered to the gelatin gel. As the bubbles grow by diffusion and thereby increase the void fraction, both amplitude and velocity of the compressive wave are reduced. Delays in the arrival of subsequent finite amplitude waves are observed by high-speed photography.

The wave propagation in a bubbly gelatin gel can be characterized with a Mach number, $Ma=\frac{1}{c} \sqrt{\Delta p/\rho_l}$, which compares the characteristic speed of the bubble collapses with the linear speed of sound in the bubbly liquid. At low void fraction and small $Ma$, the wave propagation follows Wood's~\cite{wood1941textbook} prediction for small amplitude waves. At the high void fraction regime, the linear acoustic velocity approaches the speed of collapse, i.e. $Ma$~approaches 1. In our experimental domain where a finite amplitude wave is used, the experimental results span a Ma number between about 0.2 to slightly above 1. The experimental data agree with Wood's and Campbell's models only in the asymptotic limit of small and large void fractions; the DNS simulation gives reasonable agreement over the whole measurement range. 

An important implication of this study is that the effect of the void fraction on focusing compression waves in tissue should be accounted for in medical treatments such as in histotripsy~\citep{maxwell2011}, and high-intensity focused ultrasound. While experimentally challenging, we expect that experiments in a liquid such as water should provide similar results, as predicted by our simulations.

\appendix

\section{Elastic modulus measurement}

For the measurement of elastic modulus, samples of 4, 6 and 8 percentage of gelatin by weight are prepared as described in Section~\ref{sec:exp}. The gelatin are allowed to solidify in the fridge in a cylindrical shape container with a diameter of $25.7\,$mm and height $30\,$mm. These gelatin samples are kept 72 hours in the fridge. Two hours before measurement, the samples are taken out of the fridge, and are left aside to return to room temperature. Before a measurement, a gelatin sample is removed from its container, and is placed on top of a electronic balance under a piston that is attached to a stepper motor. Both the stepper motor and the electronic balance are controlled by a home-built program from a computer. 

Our measurements are done based on the indention method as described by~\citet{hall1997phantom}. The theory behind this method is based on the fact that, uniaxial compression of an elastic material generates strain that is dependent on the material's elastic modulus, its geometry and its boundary conditions. For simplicity, we assume linear elasticity during the loading of the gelatin samples~\citep{hall1997phantom}, i.e. the gel's Young's modulus equals to the gradient of the loading portion of the stress-strain curve. The unloading portion depicts the visco-elastic property of the gel. The stress~$\sigma$~and strain~$e$~are defined as
\begin{equation}
\sigma = \frac{F(t)}{A},
\label{stresseqn}
\end{equation}

\begin{equation}
e = - \int_{h_0}^{h(t)} \frac{dh}{h},
\label{straineqn}
\end{equation}\\
where~$F(t)$~and~$A$~are the force on the sample area with area $A$ at time~$t$;~$h(t)$~is the height of the sample at time~$t$~and~$h_0$~is the initial height of the sample after the pre-loading. The sample sits on a platform that allows the sample to slip freely at its boundaries.

\subsection{Measurement parameters}

The gelatin gels are taken out of the cylindrical container and placed between the custom-made platform and the compression plate. The sample size is 13 mm in radius and 22 mm in height. The compression plat is mounted on a stage which can be mechanically moved by a stepper motor. We control the stepper motor movement from the computer using a customised software code~\citep{CDOhl2021Ramp}. The platform is placed on top of a modified kitchen scale which has been re-purposed to allow direct reading out of the measured weight at high sampling rate~\citep{virag2021repurposing}. 

\begin{figure}
\centering\includegraphics[width=0.6\linewidth]{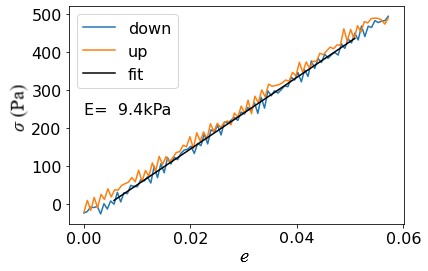}
\caption{The stress strain curve for a 6\% gelatin gel measurement. The black line is the linear estimation which is considered to be the Young's Modulus of the sample.}
\label{Fig:6PercentSressStrain}
\end{figure}

The sample is pre-loaded to -3.5 mm. Then it is subjected to 20 cycles of loading and unloading at 0.5 Hz. Ignoring the end points, the stable middle points are used to calculate the stress strain curve. A typical stress strain curve as shown in Figure~\ref{Fig:6PercentSressStrain}, from which a linear estimation is used to obtain the Young's modulus of the sample. The software code for the analysis is available from Github~\citep{CDOhl2021Analysis}. 

\subsection{Elasticity measurement results}

The Young's Modulus of the gelatin gel is depicted in Table~\ref{Table:YM}. The Young's modulus for the 4\%, 6\%, and 8\% gelatin are $6.5\pm 1.5\,$kPa, $10.7\pm 1.7 \,$kPa and $17.0\pm 0.5\,$kPa, respectively. The data is obtained using two sets of gelatine gels (each with one 4\%, one 6\%, and one 8\% gel) over three measurements. The young's modulus of the gelatin gels used in this experiment is within the stiffness range of various tissues, such as fat, skin, pancreas, kidney and liver~\citep{guimaraes2020stiffness, liu2015hydrogels}. Within this range of tissue stiffness, the wave speed measured (as indicated by the structural wave traveling through the bubbly medium), is independent of the elasticity of the medium.

\begin{table}
  \begin{center}
\def~{\hphantom{0}}
  \begin{tabular}{cc}
\hline
Gelatin Percentage (\%)&Young's Modulus (kPa)\\
\hline
4&6.5$\pm$ 1.5\\
6&10.7$\pm$ 1.7\\
8&17.0 $\pm$ 0.5\\
\hline
  \end{tabular}
  \caption{Young's Modulus of the gelatin gel measured}
  \label{Table:YM}
  \end{center}
\end{table}

\section{Image processing of the high speed photography}

\begin{figure}
\centering{
\includegraphics[width=0.4\linewidth]{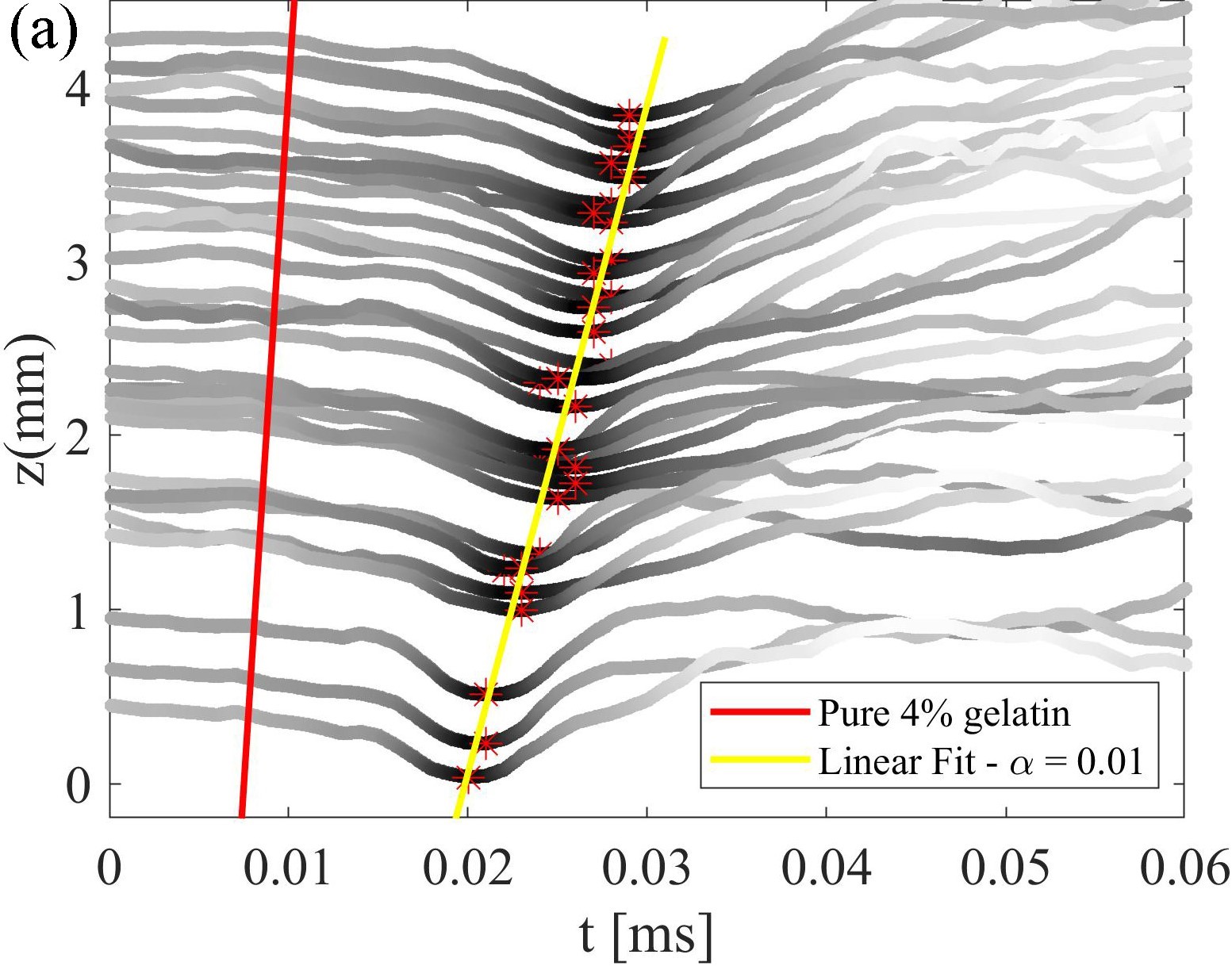}
\includegraphics[width=0.4\linewidth]{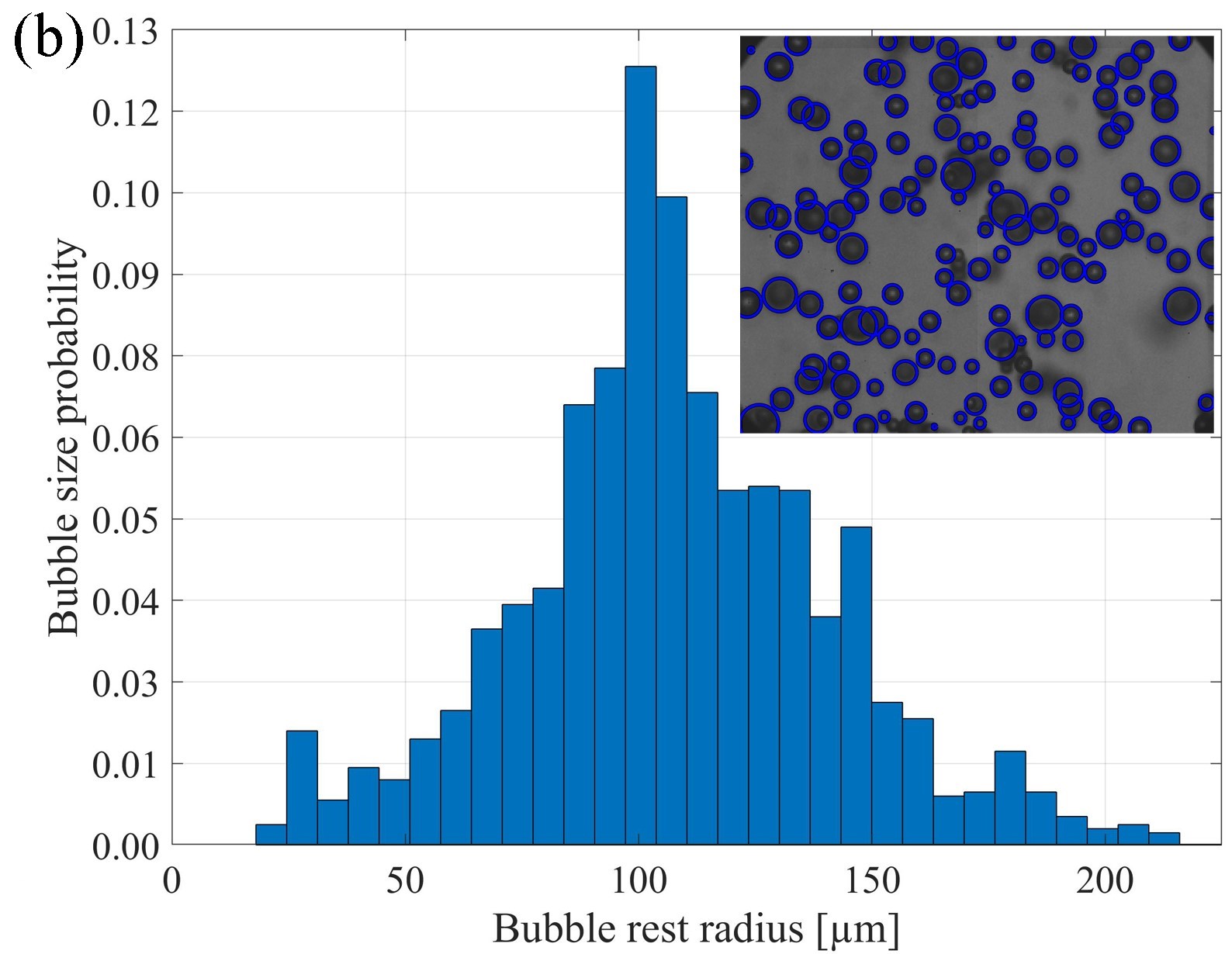}}
\caption{a) Volume dynamics of a selection of bubbles at different heights $z$ extracted from the high-speed recordings. The yellow line is connecting the minimum volume of the bubbles. Its slope is $385\,$m/s, the red line indicates a slope of $1605\,$m/s, i.e. the speed of the wave front in the pure gelatin. b) Bubble size histogram of the bubble population showed in a). The most probable bubble radius might slightly change with time due to gas diffusion into the bubbles. (Inset) Example of the bubble detection method using the Hough--transformation applied to the experimental video frames.}
\label{Fig:Bubble_Radius_Exp}
\end{figure}

The frames from the high-speed recording are analyzed with home-built algorithms to detect the bubbles, register their locations and shapes, and link them across frames to follow their dynamics. Similar codes have been used in~\citet{rossello2021demand}. As shown in the inset of Figure~\ref{Fig:Bubble_Radius_Exp}(b), the detection of the bubbles is achieved with the Hough--transformation, which allows the discrimination of individual bubbles even when they overlap each other in the images. 

The high pressure of the wave front induces a sudden collapse of gas bubbles. The wave speed was then obtained by performing a numerical fit on the position where the bubbles reach a minimum size over time using a linear  model as shown in Figure~\ref{Fig:Bubble_Radius_Exp}(a). We assume that the wave has  a constant speed in the area of  observation. This assumption of a constant speed in the short distance within the field of view of the video frames (see inset in Figure~\ref{Fig:Bubble_Radius_Exp}(b) is consistent over all 61 measurements analyzed. It is noted that as there are no bubbles in the gelatin gel initially, an alternative method was employed to obtain the wave speed for the first finite amplitude wave. Instead of the collapsing bubble, we tract the bubbles that are nucleated as the tensile phase of the finite amplitude wave passes through the gelatin gel. The same detection software was used to track the vertical coordinates of bubble nucleation to find out the distance travelled by the wave between two frames, and then the wave speed was computed with the inter-frame period of 1~$\mathrm{\mu s}$. 

The void fraction is obtained from the accumulated bubble volume assuming that the bubbles are spherical in shape. This is done within the camera frame area of 4.145 mm x 4.974 mm inscribed in the focal volume. In this small window of observation, we find a homogeneous distribution. Due to the two dimensional nature of the imaging, a projection of the bubble volume is obtained experimentally; bubbles outside the field of depth are image-blurred and are not counted. This field of depth where bubbles can be detected has been measured by translating the camera and observing the threshold of bubble detection. An error estimation for this process is used in the results for the void fraction presented in the main text. For the particular imaging setup and aperture setting of the lens used in the experiment, we find a depth of field of approximately~$5\,$mm. Thus the void fraction is calculated from $V_{gel}^{-1} \sum_i V_{b,i}$, where $V_{gel}=72.160\,$mm$^3$ and $V_{b,i}$ the volume of the $i$-th bubble in that gel volume. 

\begin{acknowledgments}
J. M. R acknowledges support by the Alexander von Humboldt Foundation (Germany) through a Georg Forster Research Fellowship. We acknowledge constructive and insightful scientific exchanges with Dr. Evert Klaseboer and Dr. Sun Qiang.
\end{acknowledgments}

\bibliography{reference}

\end{document}